\pdfoutput=1 

\documentclass[
reprint,
superscriptaddress,
 amsmath,amssymb,
 aps,
prb,
showkeys,
]{revtex4-1}

\usepackage{graphicx}
\usepackage{color}
\usepackage{dcolumn}
\usepackage{bm}
\usepackage[spanish]{babel}

\begin{document}


\title{Surface-dominated Conductivity of Few-layered Antimonene}

\author{Sahar Pakdel}
\email{pakdel.sahar@phys.au.dk}
\affiliation{Departamento de F\'isica de la Materia Condensada, Universidad Autónoma de Madrid, 28049 Madrid, Spain.}
\affiliation{School of Electrical and Computer Engineering, University College of Engineering, University of Tehran, Tehran 14395-515, Iran}
\affiliation{Department of Physics and Astronomy, Aarhus University, 8000 Aarhus C, Denmark.}

\author{J. J. Palacios}
\email{juanjose.palacios@uam.es}
\affiliation{Departamento de F\'isica de la Materia Condensada, Universidad Autónoma de Madrid, 28049 Madrid, Spain.}
\affiliation{Instituto Nicol\'as Cabrera (INC) and Condensed Matter Physics Center (IFIMAC), Universidad Autónoma de Madrid, 28049 Madrid, Spain.}
\affiliation{Department of Physics, The University of Texas at Austin, Austin, Texas 78712, USA.}



\begin{abstract}

We present a theoretical study of the phase-coherent DC conductivity of few-layered antimonene in the presence of surface disorder. It is well known that while a single layer is a trivial semiconductor, multiple layers (typically a minimum of $\approx$ 7) turn into a  semi-metal with a nontrivial topological invariant featuring protected and decoupled surface states. We employ the finite-size Kubo formalism based on density functional theory calculations to show that the conductivity is amply dominated by the topological surface states even without bulk disorder. More importantly, the conductivity of the surface states does not show traces of a metal-insulator transition while the bulk ones can be driven towards an insulating phase in presence of only surface disorder. These results suggest that few-layered antimonene, despite not being insulating in the bulk, can present many of the advantages attributed to topological insulators under very general experimental conditions.

\end{abstract}

\pacs{Valid PACS appear here}

\maketitle

\section{Introduction}
The potential of the topological surface states (TSS) in three-dimensional (3D) topological insulators (TIs) for fundamental or practical applications, remains largely unexploited due to the lack of materials meeting all the necessary requirements. 
From small bulk gap\citep{Hsieh2008} to surface degradation and band-bending issues\citep{Hsieh2009,Bianchi2010} going through unwanted bulk doping (mainly vacancies and anti-site defects)\citep{ADMA:ADMA201200187}, prototypical 3D-TIs such as Bi$_2$Te$_3$ or Bi$_2$Se$_3$  do not live up to the full theoretical expectations. Experimental studies remain mostly confined to photoelectron spectroscopy (ARPES)\citep{Hsieh2008,Chen178,Bianchi2010,Hsieh919,PhysRevB.91.161406}, although a few transport experiments have shown fingerprints of TSS in Shubnikov-de Haas oscillations\citep{Qu821,Analytis2010}, Aharonov-Bohm interference, and magnetotransport\citep{Peng2010,PhysRevLett.105.176602}. Quasiparticle interference with scanning tunneling microscopy\citep{Roushan2009,PhysRevB.87.115410} has also been explored and shows, arguably, the most clear signatures of the topological nature of the surface states. 

Ideally, if the Dirac point lies in the middle of the bulk gap and the Fermi energy can be tuned in its vicinity, the conductivity of one surface is expected to be ``universal'' ($\approx e^2/h$) and exhibit anti-localization regardless of disorder strength\citep{Geim2007}. However, experimental verification of this behavior faces a major difficulty since the surface/bulk conductivity ratio is typically a small fraction and separating the 2D conductivity from the often unavoidable and dominating 3D bulk contribution is challenging\citep{acs.nanolett.5b04425}. Finding materials with a large surface/bulk conductivity ratio is therefore, of much current interest. For instance, research on ternary compound TIs such as Bi$_2$Te$_{2}$Se\citep{PhysRevB.84.235206,PhysRevB.88.081301,PhysRevB.90.165140}, Bi$_2$Te$_{2.4}$Se$_{0.6}$\citep{PhysRevB.89.125416} or (Bi$_{1-x}$Sb$_{x}$)$_2$Te$_3$\citep{Zhang2011,Kong2011,0953-8984-28-49-495501}, and even quaternary  compounds such as Bi$_{2-x}$Sb$_x$Te$_{3-y}$Se$_y$\citep{Arakane2012} or Sn-doped Bi$_{1.1}$Sb$_{0.9}$Te$_2$S\citep{Kushwaha2016}, typically in thin film form, has demonstrated enhanced and tunable Dirac surface contribution over bulk conduction. 

As the complexity of the material composition grows, the chances for practical use, however, decrease. An elemental bulk material such as Sb happens to be a topological semi-metal due to it's inverted bulk band order\citep{PhysRevB.76.045302}.
Despite the absence of a bulk gap, it's topological invariant guarantees that antimony features protected TSS, coexisting with bulk bands at the Fermi energy\citep{Hsieh919,PhysRevLett.107.036802,PhysRevB.91.161406}. 
Thin films of Sb(111) could in principle become a 3D-TI if quantum confinement opened a gap in the bulk bands. However, when thin enough, coupling between the TSS on opposite surfaces also occurs and this is expected to degrade or even destroy the TSS exotic properties such as their expected protection against backscattering.
Ultimately, a single Sb(111) layer (or monolayer antimonene) becomes a trivial semiconductor\citep{ANIE:ANIE201411246}. (Note that the term monolayer antimonene used in this work is sometimes referred to as bilayer Sb(111) in the literature.) Recently, it has been shown that few-layered (FL) antimonene and even monolayer antimonene can be obtained through mechanical exfoliation techniques\citep{ADMA:ADMA201602128}. Importantly, antimonene, in contrast to other elemental 2D crystals such as the popular phosphorene\citep{phosphorene}, happens to be surprisingly robust under ambient conditions.

Theoretical studies on FL antimonene have shown that the decoupling of the TSS requires a minimum of $\simeq$ 7 layers\citep{PhysRevB.85.201410,PhysRevLett.108.176401}.
In between the semiconductor monolayer and the 7-layered antimonene a crossover occurs where claims on the existence of a 2D topological insulator have also been reported\citep{PhysRevB.85.201410}. When the TSS of opposite surfaces are decoupled and the gap at the Dirac point is closed down, the Fermi energy crosses the Dirac cone above the Dirac point, but also crosses 6 hole pockets and 3 electron pockets (see, e.g., Ref. \onlinecite{Hsieh919} and Fig. \ref{bands}). For a number of layers below $\approx 50$, they all result from a single 2D band. While the electron pockets have a bulk character, the hole pockets are partially formed by helical surface and trivial bulk states. 

Concerning fundamental transport properties, it is known that materials in the 2D symplectic class present a metal-insulator transition (MIT) with a critical conductivity (for some simplified models) of $\sigma_c \approx 1.42$ e$^2$/h\citep{0305-4470-39-13-003,PhysRevLett.89.256601}. There, according to the single-parameter scaling theory of localization\cite{PhysRevLett.42.673}, the $\beta$ function changes sign. On the other hand, the $\beta$ function is known to be always positive for an odd number of Dirac cones\citep{PhysRevLett.99.106801}. In our case,  carriers belonging to the bulk electron pockets near M (see Fig. \ref{bands}) should certainly exhibit a MIT. However, concerning the states around $\Gamma$, the situation is less clear. While along the $\Gamma$-K direction only one type of carriers (those belonging to the central Dirac cone) is present at the Fermi level, along the $\Gamma$-M direction, the band structure resembles more that of 2D free electrons with Rashba coupling. In addition, bulk states are also present along this direction. 
This complexity triggers a number of possibilities, namely, if the TSS do not present a MIT, the conductivity will be dominated by them for sufficiently strong disorder (which will necessarily eliminate bulk contribution). On the other hand, if the TSS present a MIT, the question is whether there is a unique critical disorder for both bulk states and TSS or either type of carrier can be found on different sides of their respective critical points. In our calculations no traces of a MIT for the TSS appear and, although we cannot give a definite answer to this question due to computational limitations, we show that the TSS and bulk states do not share the same critical disorder. Remarkably, this fact and, in general, the dominant surface character of the conductivity occur considering a perfect crystal with only surface disorder.

From a modeling point of view, effective Hamiltonians can account for the major features of the band structure near the Dirac point. However, the complexity of the actual bands in most common 3D-TIs (and also in FL antimonene) comes from the fact that the Fermi level rarely lies close to the Dirac point. Moreover, the Dirac point does not necessarily lie in the bulk gap, typically lying close or overlapping with the bulk valence band or the surface bands themselves\citep{Bianchi2010}. As far as conductivity is concerned, existing theoretical work has not addressed these more realistic situations.  In order to address the questions raised in the previous paragraph, here we present conductivity calculations for FL-antimonene using the Kubo formalism in its numerical finite-size version, calculated upon the band structure and wave functions obtained from density functional theory (DFT).

 The typical medium cluster approach (and previous methodological attempts) aims at a full description of Anderson localization in real systems where disorder and electron-electron interactions are treated on equal footing (see for e.g. the review by Terletska et al.\citep{terletska2018systematic} and references therein). In this regard, our reason to start from a DFT description of antimonene is not based on any expectation that interactions could play a significant role, but more on the need for an accurate and realistic single-particle Hamiltonian to work with. There is no doubt on the interest of these type of studies, but we cannot perform self-consistency after introducing disorder.  Note that our focus here is surface disorder and the topologically protected surface states derive from a clean bulk system. The topological origin of the surface states requires a clean bulk and these cannot be isolated and independently  described from the bulk.

\section{Numerical method description}
\begin{figure}[h]
\includegraphics[scale=0.82]{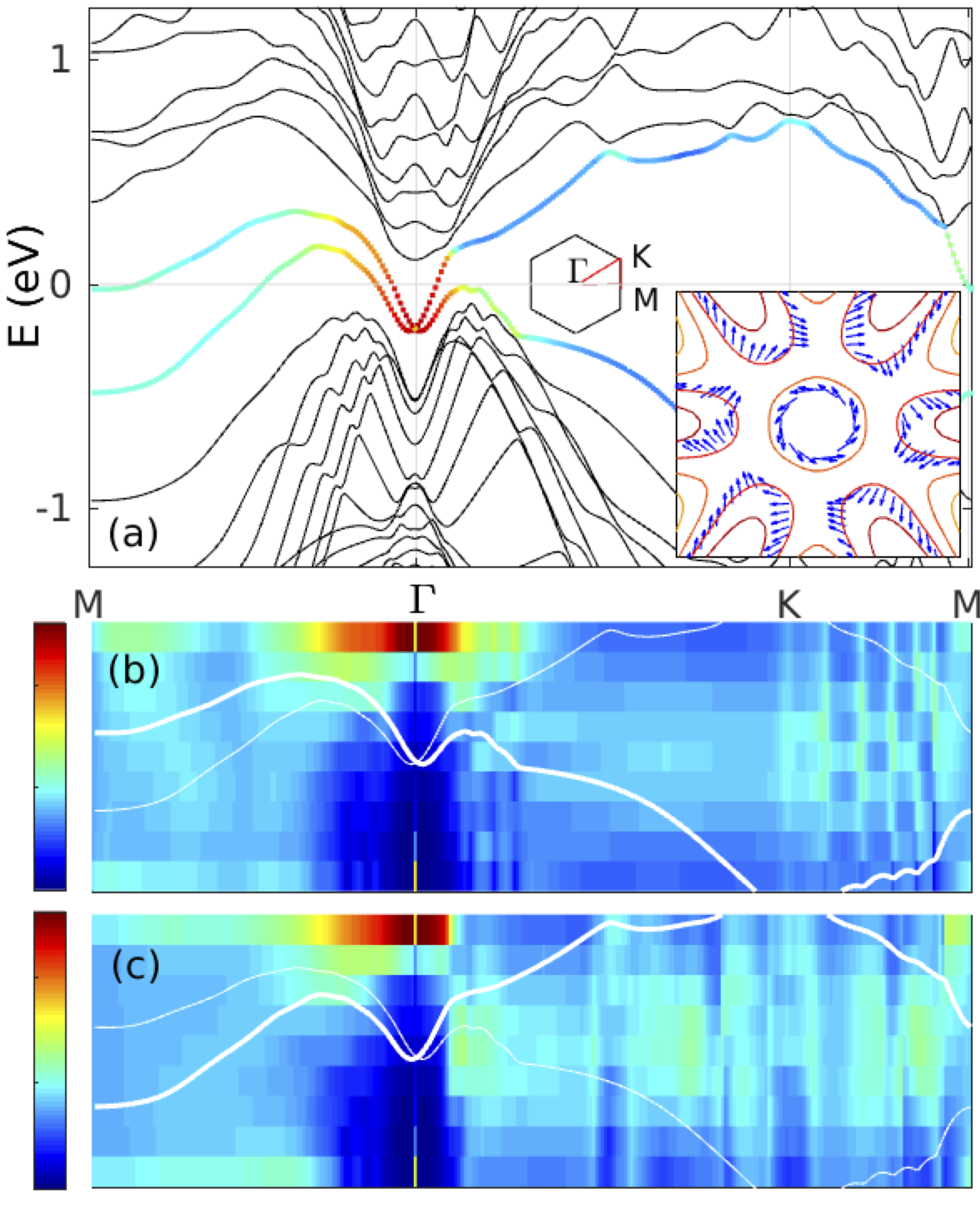}
\caption{Color online. (a) Electronic band structure of 9-layers of antimonene. Color code indicates the surface (red) or bulk (dark blue) character of the Bloch states for the bands closer to the Fermi energy. (b) and (c) Charge density distribution of the Bloch states (denoted by the thick white lines) across the multilayer (different Sb layers in the vertical axes). Color code indicates red for high density and dark blue for low density values. We are only plotting one copy of degenerate bands. }
\label{bands}
\end{figure}

 We start by showing in Fig. \ref{bands}(a) the DFT band structure of 9 layers of Sb. We have used the CRYSTAL code\citep{dovesi2014crystal14,dovesicrystal14,crystal} to obtain the Kohn-Sham Hamiltonian to which we have added SOC after selfconsistency\citep{pakdel2018implementation}. The two colored bands crossing the Fermi energy are the relevant ones for conductivity calculations. In Fig. \ref{bands}(b) and (c) we plot the layer-by-layer charge density  of the Bloch states associated with these bands. Near $\Gamma$ the states are confined to one surface (the top one in this case since only one of the doubly degenerate bands is shown in each panel). Away from $\Gamma$ the charge density spreads across the multilayer, presenting a strong bulk character along $\Gamma-$K, but a less bulk character along $\Gamma-$M. In the latter direction a hole pocket is observed where the states transit from having a strong top-surface character near $\Gamma$ to being localized on both surfaces [see Fig. \ref{bands}(c)]. Inset in Fig. \ref{bands}(a) shows the Fermi surface and the spin character of the Bloch states. The helical counter-clockwise behavior around the $\Gamma$ point is evident along with the helical clockwise character of the hole pocket states close to the Dirac cone. These results fully agree with those in the literature and manifest the topological character of FL antimonene. 

To evaluate the zero-bias conductivity we use the finite-size Kubo formalism\citep{PhysRevLett.98.076602}:
\begin{equation}
\sigma_{xx}=\dfrac{-i\hbar e^2}{L^2}\sum\limits_{i,j}
\dfrac{f(E_i)-f(E_j)}{E_i-E_{j}}
\dfrac{\langle \chi_i\vert \hat v_x \vert \chi_j\rangle \langle \chi_j\vert \hat v_x \vert \chi_i\rangle}{E_i-E_{j}+i\eta},
\label{kubo}
\end{equation}

where $\vert \chi_{i}\rangle$ and $E_i$ represent the eigen states and energies of the crystal Hamiltonian plus disorder, respectively. $f(E)$ is the Fermi-Dirac distribution function and $L$ denotes the size of the system. We have studied a short-range scattering model for delta impurities assuming that it does not mix intra-atomic orbitals, as defined below:

\begin{equation*}
\resizebox{1.0\hsize}{!}
{$
\langle n,k \vert V_{dis} \vert m,k'\rangle=\dfrac{1}{N}
\sum\limits_{imp}\sum\limits_{\alpha}
V_lC_{nl\alpha}^{*}(k)C_{ml\alpha}(k') e^{-i(k-k')R_{\rm imp}},
$}
\end{equation*}

where $\vert  n,k \rangle$ represents a Bloch state in the $n-th$ band with wave vector $k$ and energy $\epsilon_{n}(k)$. The strength of disorder potential, $V_l$, is chosen to be constant with random signs and locations for $N$ impurities at Bravais lattice  vectors $R_{\rm imp}$ and affecting the $l$-th sublattice. Bloch states are defined by the coefficients $C_{nl\alpha}(k)$ where $\alpha$ denotes the atomic basis set element. The impurities are placed on both surfaces with distinct random distributions. Vectors are implied throughout the text.

Since our lattice consists of a multi-atomic unit cell with multi-orbital atoms, the evaluation of the velocity matrix elements in Eq. \ref{kubo} needs to be carried out with some care. A proper gauge choice simplifies  this evaluation:
\begin{eqnarray*}
\langle n,k \vert \hat{v} \vert m,k \rangle \approx \dfrac{1}{\hbar} \sum\limits_{ij} \sum\limits_{\alpha\beta} C_{ni\alpha}^{*}(k)C_{mj\beta}(k)   
\nabla_k H^{ij}_{\alpha\beta}(k)  \nonumber \\
+\dfrac{i}{\hbar} \sum\limits_{\alpha\beta} C_{ni\alpha}(k)^{*}C_{mi\beta}(k) [\epsilon_n(k)-\epsilon_m(k)] D_{\beta\alpha},
\end{eqnarray*}
where $D_{\beta\alpha}=\langle R,i,\alpha \vert r\vert R,i,\beta\rangle$ is the dipolar term between different atomic orbitals at the same atom\citep{pedersen2001optical}. 

\begin{figure}[h]
\includegraphics[scale=0.51]{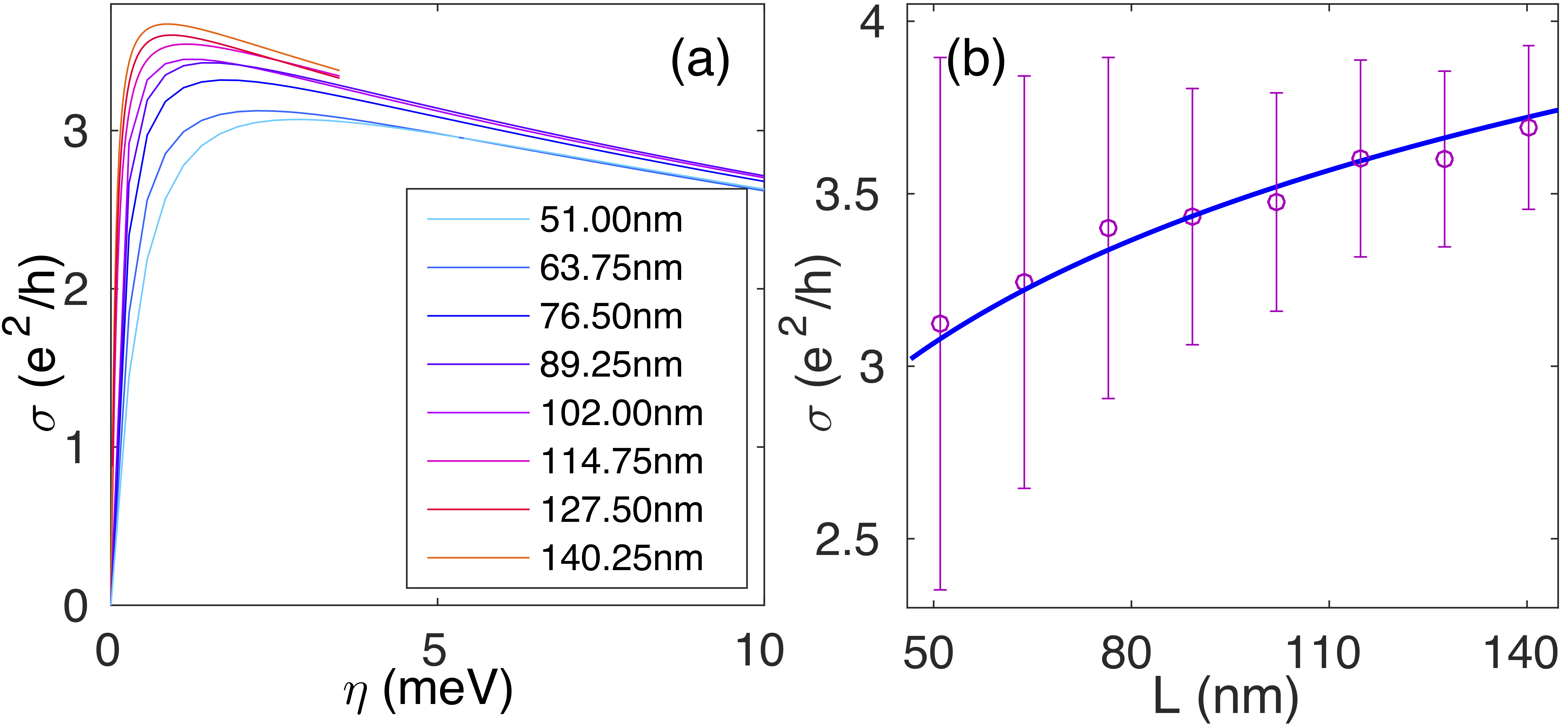}
\caption{Color online. (a) Kubo formula conductivities of the Dirac cone at the neutrality point as a function of $\eta$ and for different sizes of the system ($L$). (b) Maximum of the conductivity curves in (a) as a function of $L$. Logarithmic curve has been fitted to the data of panel (b).}
\label{Dirac}
\end{figure}

\section{Results and Discussion}
We start by evaluating the conductivity associated with the Dirac electrons of this system, which will also serve as a verification of our implementation. Following Ref. \onlinecite{PhysRevLett.98.076602}, Fig. \ref{Dirac} shows $\sigma$ as a function of the imaginary part in Eq. \ref{kubo} for different sample sizes $L$, and a given disorder strength and concentration ($\delta$,$n_{i}$)=(30, 2.3$\%$). All curves show weak dependence on $\eta$ in a finite range and reach a maximum before dropping to zero for very small values. We take this maximum of $\sigma$ vs $\eta$ as the actual conductivity value. As expected, the conductivity shows the anticipated smooth logarithmic behavior characteristic of Dirac electrons fitted by the equation $\sigma \approx 0.54 {\rm Ln}(L/\ell)$ [Fig. \ref{Dirac}(b)].

\begin{figure}[h]
\includegraphics[trim=22 20 0 0, scale=0.57]{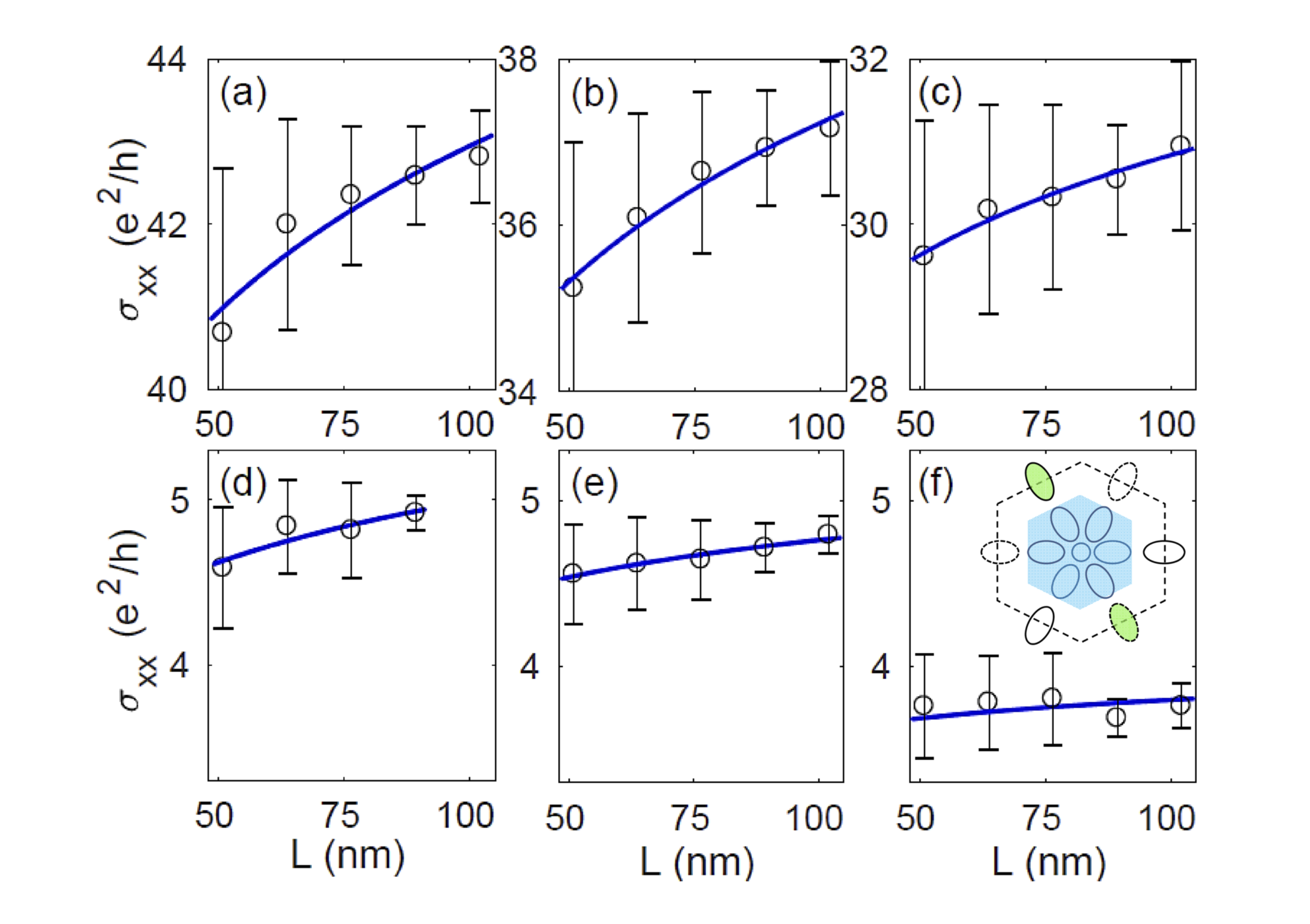}
\caption{Color online. $\sigma_{xx}$ associated with TSS (a-c) and one bulk pocket (d-f) as a function of the size of the system, for different disorder strengths and concentrations (50,3.2$\%$), (60,3.2$\%$) and (60,3.7$\%$) from left to right columns. Logarithmic curves fitted to the data are shown in blue in all panels. The schematic inset in (f) shows the BZ regions considered for TSS and the chosen bulk pocket with blue and green shades, respectively. }
\label{scaling}
\end{figure}

We now compute, separately, the conductivity of the Dirac cone plus the six surrounding hole pockets on one hand and one of the electron pockets on the other. The regions associated with these two types of states in the Brillouin zone (BZ) are specified in the inset of Fig. \ref{scaling}(f) with blue and green shades, respectively. We first consider the case of weak disorder assuming that disorder does not mix bulk and TSS pockets as it would be certainly the case for sufficiently long-ranged disorder (our conclusions are not conditioned by the type of disorder we have considered).  The behavior of $\sigma_{xx}$ for different weak disorder cases, as shown in Figs. \ref{scaling}(a-c), is the expected one for the 2D electrons in the symplectic class, namely, the conductivity increases with the size of the system with overall high values of conductivity (both surfaces are taken into account here). Increasing either the strength or the concentration of disorder, both the magnitude and the size-dependence of the conductivity decrease, but the scaling remains the same. 

\begin{figure}[h]
\includegraphics[trim=0 20 0 0,scale=0.64]{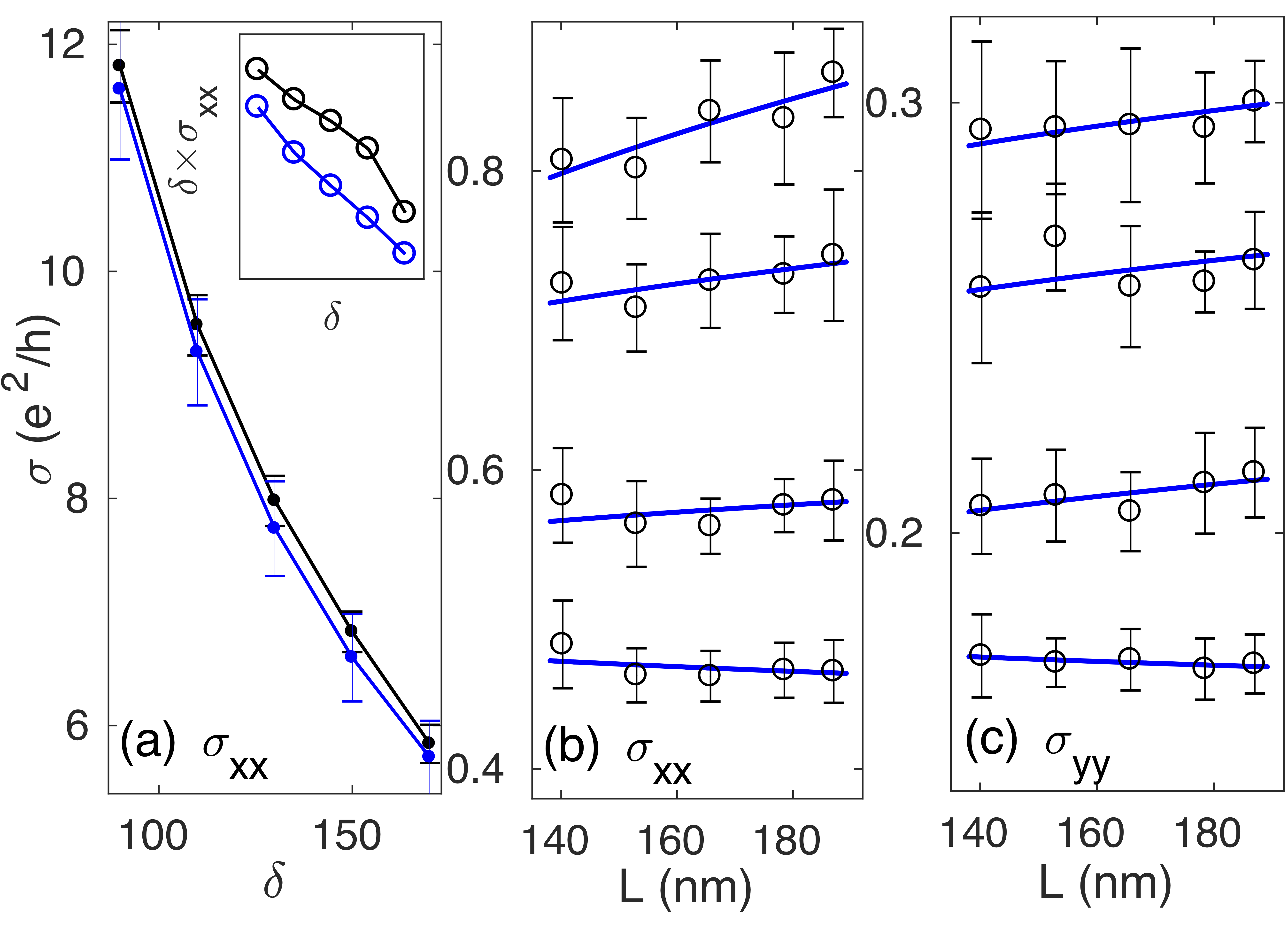}
\caption{(a) $\sigma_{xx}$ of TSS for $L=51$nm and 102 nm as a function of disorder strength ($\delta$) in black and blue curves respectively. The Inset presents the same results plotting $\sigma_{xx}\times\delta$ as a function of $\delta$. (b) and (c) panels show $\sigma_{xx}$ and $\sigma_{yy}$ for bulk states (shown with green shade in the inset of Fig. \ref{scaling}(f)) as a function of $L$ for $\delta$= 120, 130, 150, 170 from top to bottom. Logarithmic curves fitted to the data of (b) and (c). $n_i$=3.2$\%$ is considered in all cases. }
\label{MIT}
\end{figure}

Assuming again that scattering is absent between far away pockets in the BZ, we compute the conductivity of just one of the bulk pockets [the results are shown Figs. \ref{scaling}(d-f)]. The behavior with size is similar to that of the TSS, also scaling to a metal. 
 Working with only one pocket results in an anisotropic conductivity, with $\sigma_{xx}$ larger than $\sigma_{yy}$ in this case (see Fig. \ref{MIT}), while the total bulk conductivity should be, of course, isotropic. We have verified that this is the case with a few disorder realizations summing over the three bulk pockets, but we can estimate an upper limit to the total conductivity by simply multiplying the average $(\sigma_{xx}+\sigma_{yy})/2$ by three. Even so, the main contribution to the total conductivity is predominantly given by the TSS up to the highest values of disorder considered. The overall smaller values of the bulk conductivity with respect to the TSS  ones are somewhat unexpected since we are only considering surface disorder and a dominant contribution of bulk bands could have been naively expected.

We now turn our attention to strong disorder, chosen in a way that brings the conductivity closer to the quantum of conductance where a metal-insulator transition (MIT) may be expected.  Panels (b) and (c) of Fig. \ref{MIT} show $\sigma_{xx}$ and $\sigma_{yy}$ of the bulk pocket shown in the inset of Fig. \ref{scaling}(f) with green shade. The behavior of the conductivity in different directions is similar, but it is less pronounced in $\sigma_{yy}$  because of the smaller overall values. As we increase the disorder strength from top to bottom, the three upper curves clearly increase with size indicating metallic behavior, while the bottom one starts to show a decreasing trend expected for an insulator. The transition seems to occur in both panels at the same disorder strength (approximately between the third and forth curves), although the critical conductivities logically differ from each other.

Since the calculations of the conductivity of TSS are computationally more demanding and we have to average over many disordered systems, we choose to study the scaling behavior of the TSS in this limit of disorder differently. 
In Fig. \ref{MIT}(a) we show the conductivity of TSS for the smallest and largest sizes of the systems studied in panels (b-c), as a function of the disorder strength. The largest disorder in all three panels is chosen to be the same. In panel (a) we see that both curves run essentially parallel to each other with no indication of a crossing  at the suggested critical disorder value for bulk states ($\delta \approx 150$). Due to computational limitations we cannot report larger system sizes which could show the separation of the two curves more clearly. However, the inset of \ref{MIT}(a), showing the conductivity times disorder strength shows that the two curves do not cross in the studied disorder range. Notice, though, that we cannot entirely discard a MIT at higher disorder strengths, but, if it exists, it will happen at a critical value larger than that for bulk states.

\section{Conclusion}
In conclusion, we have shown that the conductivity of surface disordered FL antimonene is dominated by the TSS with the bulk pockets playing a less significant role. This happens in two scenarios. For weak disorder, the conductivity of both bulk and TSS carriers scales towards a metal with sample sizes. On the other hand, for strong disorder, bulk conductivity moves towards a MIT transition, as expected for electrons in the symplectic class, while the conductivity of the TSS still scales to a metal for similar disorder values. This result is even more remarkable considering that we have only considered surface disorder, which is the most relevant case for a possible experimental verification. Our  results  open a venue for gapless semi-metallic topological materials to serve experimental purposes usually reserved to topological insulators.

\section{Acknowledgments}
JJP acknowledges enlightening conversations with A. H. MacDonald and also thanks F. Zamora and J. G\'omez-Herrero for sharing preliminary experimental results. This work has been supported by funds from Spanish MINECO through Grant FIS2016-80434-P, the Fundaci\'on Ram\'on Areces, the Mar\' ia de Maeztu Program for Units of Excellence in R$\&$D (MDM-2014-0377), the Comunidad Autonoma de Madrid through S2018/NMT-4321 (NanomagCOST-CM) and the European Union Seventh Framework Programme under Grant agreement No. 604391 Graphene Flagship. SP acknowledges the computer resources and assistance provided by the Centro de Computación Científica of the Universidad Autonoma de Madrid and the RES. SP was also supported by the VILLUM FONDEN via the Centre of Excellence for Dirac Materials (Grant No. 11744)


\begin{thebibliography}{42}%
\makeatletter
\providecommand \@ifxundefined [1]{%
 \@ifx{#1\undefined}
}%
\providecommand \@ifnum [1]{%
 \ifnum #1\expandafter \@firstoftwo
 \else \expandafter \@secondoftwo
 \fi
}%
\providecommand \@ifx [1]{%
 \ifx #1\expandafter \@firstoftwo
 \else \expandafter \@secondoftwo
 \fi
}%
\providecommand \natexlab [1]{#1}%
\providecommand \enquote  [1]{``#1''}%
\providecommand \bibnamefont  [1]{#1}%
\providecommand \bibfnamefont [1]{#1}%
\providecommand \citenamefont [1]{#1}%
\providecommand \href@noop [0]{\@secondoftwo}%
\providecommand \href [0]{\begingroup \@sanitize@url \@href}%
\providecommand \@href[1]{\@@startlink{#1}\@@href}%
\providecommand \@@href[1]{\endgroup#1\@@endlink}%
\providecommand \@sanitize@url [0]{\catcode `\\12\catcode `\$12\catcode
  `\&12\catcode `\#12\catcode `\^12\catcode `\_12\catcode `\%12\relax}%
\providecommand \@@startlink[1]{}%
\providecommand \@@endlink[0]{}%
\providecommand \url  [0]{\begingroup\@sanitize@url \@url }%
\providecommand \@url [1]{\endgroup\@href {#1}{\urlprefix }}%
\providecommand \urlprefix  [0]{URL }%
\providecommand \Eprint [0]{\href }%
\providecommand \doibase [0]{http://dx.doi.org/}%
\providecommand \selectlanguage [0]{\@gobble}%
\providecommand \bibinfo  [0]{\@secondoftwo}%
\providecommand \bibfield  [0]{\@secondoftwo}%
\providecommand \translation [1]{[#1]}%
\providecommand \BibitemOpen [0]{}%
\providecommand \bibitemStop [0]{}%
\providecommand \bibitemNoStop [0]{.\EOS\space}%
\providecommand \EOS [0]{\spacefactor3000\relax}%
\providecommand \BibitemShut  [1]{\csname bibitem#1\endcsname}%
\let\auto@bib@innerbib\@empty
\bibitem [{\citenamefont {Hsieh}\ \emph {et~al.}(2008)\citenamefont {Hsieh},
  \citenamefont {Qian}, \citenamefont {Wray}, \citenamefont {Xia},
  \citenamefont {Hor}, \citenamefont {Cava},\ and\ \citenamefont
  {Hasan}}]{Hsieh2008}%
  \BibitemOpen
  \bibfield  {author} {\bibinfo {author} {\bibfnamefont {D.}~\bibnamefont
  {Hsieh}}, \bibinfo {author} {\bibfnamefont {D.}~\bibnamefont {Qian}},
  \bibinfo {author} {\bibfnamefont {L.}~\bibnamefont {Wray}}, \bibinfo {author}
  {\bibfnamefont {Y.}~\bibnamefont {Xia}}, \bibinfo {author} {\bibfnamefont
  {Y.~S.}\ \bibnamefont {Hor}}, \bibinfo {author} {\bibfnamefont {R.~J.}\
  \bibnamefont {Cava}}, \ and\ \bibinfo {author} {\bibfnamefont {M.~Z.}\
  \bibnamefont {Hasan}},\ }\href@noop {} {\bibfield  {journal} {\bibinfo
  {journal} {Nature}\ }\textbf {\bibinfo {volume} {452}},\ \bibinfo {pages}
  {970} (\bibinfo {year} {2008})}\BibitemShut {NoStop}%
\bibitem [{\citenamefont {Hsieh}\ \emph
  {et~al.}(2009{\natexlab{a}})\citenamefont {Hsieh}, \citenamefont {Xia},
  \citenamefont {Qian}, \citenamefont {Wray}, \citenamefont {Dil},
  \citenamefont {Meier}, \citenamefont {Osterwalder}, \citenamefont {Patthey},
  \citenamefont {Checkelsky}, \citenamefont {Ong}, \citenamefont {Fedorov},
  \citenamefont {Lin}, \citenamefont {Bansil}, \citenamefont {Grauer},
  \citenamefont {Hor}, \citenamefont {Cava},\ and\ \citenamefont
  {Hasan}}]{Hsieh2009}%
  \BibitemOpen
  \bibfield  {author} {\bibinfo {author} {\bibfnamefont {D.}~\bibnamefont
  {Hsieh}}, \bibinfo {author} {\bibfnamefont {Y.}~\bibnamefont {Xia}}, \bibinfo
  {author} {\bibfnamefont {D.}~\bibnamefont {Qian}}, \bibinfo {author}
  {\bibfnamefont {L.}~\bibnamefont {Wray}}, \bibinfo {author} {\bibfnamefont
  {J.~H.}\ \bibnamefont {Dil}}, \bibinfo {author} {\bibfnamefont
  {F.}~\bibnamefont {Meier}}, \bibinfo {author} {\bibfnamefont
  {J.}~\bibnamefont {Osterwalder}}, \bibinfo {author} {\bibfnamefont
  {L.}~\bibnamefont {Patthey}}, \bibinfo {author} {\bibfnamefont {J.~G.}\
  \bibnamefont {Checkelsky}}, \bibinfo {author} {\bibfnamefont {N.~P.}\
  \bibnamefont {Ong}}, \bibinfo {author} {\bibfnamefont {A.~V.}\ \bibnamefont
  {Fedorov}}, \bibinfo {author} {\bibfnamefont {H.}~\bibnamefont {Lin}},
  \bibinfo {author} {\bibfnamefont {A.}~\bibnamefont {Bansil}}, \bibinfo
  {author} {\bibfnamefont {D.}~\bibnamefont {Grauer}}, \bibinfo {author}
  {\bibfnamefont {Y.~S.}\ \bibnamefont {Hor}}, \bibinfo {author} {\bibfnamefont
  {R.~J.}\ \bibnamefont {Cava}}, \ and\ \bibinfo {author} {\bibfnamefont
  {M.~Z.}\ \bibnamefont {Hasan}},\ }\href@noop {} {\bibfield  {journal}
  {\bibinfo  {journal} {Nature}\ }\textbf {\bibinfo {volume} {460}},\ \bibinfo
  {pages} {1101} (\bibinfo {year} {2009}{\natexlab{a}})}\BibitemShut {NoStop}%
\bibitem [{\citenamefont {Bianchi}\ \emph {et~al.}(2010)\citenamefont
  {Bianchi}, \citenamefont {Guan}, \citenamefont {Bao}, \citenamefont {Mi},
  \citenamefont {Iversen}, \citenamefont {King},\ and\ \citenamefont
  {Hofmann}}]{Bianchi2010}%
  \BibitemOpen
  \bibfield  {author} {\bibinfo {author} {\bibfnamefont {M.}~\bibnamefont
  {Bianchi}}, \bibinfo {author} {\bibfnamefont {D.}~\bibnamefont {Guan}},
  \bibinfo {author} {\bibfnamefont {S.}~\bibnamefont {Bao}}, \bibinfo {author}
  {\bibfnamefont {J.}~\bibnamefont {Mi}}, \bibinfo {author} {\bibfnamefont
  {B.~B.}\ \bibnamefont {Iversen}}, \bibinfo {author} {\bibfnamefont
  {P.~D.~C.}\ \bibnamefont {King}}, \ and\ \bibinfo {author} {\bibfnamefont
  {P.}~\bibnamefont {Hofmann}},\ }\href@noop {} {\bibfield  {journal} {\bibinfo
   {journal} {Nat. Commun.}\ }\textbf {\bibinfo {volume} {1}},\ \bibinfo
  {pages} {128} (\bibinfo {year} {2010})}\BibitemShut {NoStop}%
\bibitem [{\citenamefont {Scanlon}\ \emph {et~al.}(2012)\citenamefont
  {Scanlon}, \citenamefont {King}, \citenamefont {Singh}, \citenamefont {de~la
  Torre}, \citenamefont {Walker}, \citenamefont {Balakrishnan}, \citenamefont
  {Baumberger},\ and\ \citenamefont {Catlow}}]{ADMA:ADMA201200187}%
  \BibitemOpen
  \bibfield  {author} {\bibinfo {author} {\bibfnamefont {D.~O.}\ \bibnamefont
  {Scanlon}}, \bibinfo {author} {\bibfnamefont {P.~D.~C.}\ \bibnamefont
  {King}}, \bibinfo {author} {\bibfnamefont {R.~P.}\ \bibnamefont {Singh}},
  \bibinfo {author} {\bibfnamefont {A.}~\bibnamefont {de~la Torre}}, \bibinfo
  {author} {\bibfnamefont {S.~M.}\ \bibnamefont {Walker}}, \bibinfo {author}
  {\bibfnamefont {G.}~\bibnamefont {Balakrishnan}}, \bibinfo {author}
  {\bibfnamefont {F.}~\bibnamefont {Baumberger}}, \ and\ \bibinfo {author}
  {\bibfnamefont {C.~R.~A.}\ \bibnamefont {Catlow}},\ }\href@noop {} {\bibfield
   {journal} {\bibinfo  {journal} {Adv. Mater.}\ }\textbf {\bibinfo {volume}
  {24}},\ \bibinfo {pages} {2154} (\bibinfo {year} {2012})}\BibitemShut
  {NoStop}%
\bibitem [{\citenamefont {Chen}\ \emph {et~al.}(2009)\citenamefont {Chen},
  \citenamefont {Analytis}, \citenamefont {Chu}, \citenamefont {Liu},
  \citenamefont {Mo}, \citenamefont {Qi}, \citenamefont {Zhang}, \citenamefont
  {Lu}, \citenamefont {Dai}, \citenamefont {Fang}, \citenamefont {Zhang},
  \citenamefont {Fisher}, \citenamefont {Hussain},\ and\ \citenamefont
  {Shen}}]{Chen178}%
  \BibitemOpen
  \bibfield  {author} {\bibinfo {author} {\bibfnamefont {Y.~L.}\ \bibnamefont
  {Chen}}, \bibinfo {author} {\bibfnamefont {J.~G.}\ \bibnamefont {Analytis}},
  \bibinfo {author} {\bibfnamefont {J.-H.}\ \bibnamefont {Chu}}, \bibinfo
  {author} {\bibfnamefont {Z.~K.}\ \bibnamefont {Liu}}, \bibinfo {author}
  {\bibfnamefont {S.-K.}\ \bibnamefont {Mo}}, \bibinfo {author} {\bibfnamefont
  {X.~L.}\ \bibnamefont {Qi}}, \bibinfo {author} {\bibfnamefont {H.~J.}\
  \bibnamefont {Zhang}}, \bibinfo {author} {\bibfnamefont {D.~H.}\ \bibnamefont
  {Lu}}, \bibinfo {author} {\bibfnamefont {X.}~\bibnamefont {Dai}}, \bibinfo
  {author} {\bibfnamefont {Z.}~\bibnamefont {Fang}}, \bibinfo {author}
  {\bibfnamefont {S.~C.}\ \bibnamefont {Zhang}}, \bibinfo {author}
  {\bibfnamefont {I.~R.}\ \bibnamefont {Fisher}}, \bibinfo {author}
  {\bibfnamefont {Z.}~\bibnamefont {Hussain}}, \ and\ \bibinfo {author}
  {\bibfnamefont {Z.-X.}\ \bibnamefont {Shen}},\ }\href@noop {} {\bibfield
  {journal} {\bibinfo  {journal} {Science}\ }\textbf {\bibinfo {volume}
  {325}},\ \bibinfo {pages} {178} (\bibinfo {year} {2009})}\BibitemShut
  {NoStop}%
\bibitem [{\citenamefont {Hsieh}\ \emph
  {et~al.}(2009{\natexlab{b}})\citenamefont {Hsieh}, \citenamefont {Xia},
  \citenamefont {Wray}, \citenamefont {Qian}, \citenamefont {Pal},
  \citenamefont {Dil}, \citenamefont {Osterwalder}, \citenamefont {Meier},
  \citenamefont {Bihlmayer}, \citenamefont {Kane}, \citenamefont {Hor},
  \citenamefont {Cava},\ and\ \citenamefont {Hasan}}]{Hsieh919}%
  \BibitemOpen
  \bibfield  {author} {\bibinfo {author} {\bibfnamefont {D.}~\bibnamefont
  {Hsieh}}, \bibinfo {author} {\bibfnamefont {Y.}~\bibnamefont {Xia}}, \bibinfo
  {author} {\bibfnamefont {L.}~\bibnamefont {Wray}}, \bibinfo {author}
  {\bibfnamefont {D.}~\bibnamefont {Qian}}, \bibinfo {author} {\bibfnamefont
  {A.}~\bibnamefont {Pal}}, \bibinfo {author} {\bibfnamefont {J.~H.}\
  \bibnamefont {Dil}}, \bibinfo {author} {\bibfnamefont {J.}~\bibnamefont
  {Osterwalder}}, \bibinfo {author} {\bibfnamefont {F.}~\bibnamefont {Meier}},
  \bibinfo {author} {\bibfnamefont {G.}~\bibnamefont {Bihlmayer}}, \bibinfo
  {author} {\bibfnamefont {C.~L.}\ \bibnamefont {Kane}}, \bibinfo {author}
  {\bibfnamefont {Y.~S.}\ \bibnamefont {Hor}}, \bibinfo {author} {\bibfnamefont
  {R.~J.}\ \bibnamefont {Cava}}, \ and\ \bibinfo {author} {\bibfnamefont
  {M.~Z.}\ \bibnamefont {Hasan}},\ }\href@noop {} {\bibfield  {journal}
  {\bibinfo  {journal} {Science}\ }\textbf {\bibinfo {volume} {323}},\ \bibinfo
  {pages} {919} (\bibinfo {year} {2009}{\natexlab{b}})}\BibitemShut {NoStop}%
\bibitem [{\citenamefont {Benia}\ \emph {et~al.}(2015)\citenamefont {Benia},
  \citenamefont {Stra\ss{}er}, \citenamefont {Kern},\ and\ \citenamefont
  {Ast}}]{PhysRevB.91.161406}%
  \BibitemOpen
  \bibfield  {author} {\bibinfo {author} {\bibfnamefont {H.~M.}\ \bibnamefont
  {Benia}}, \bibinfo {author} {\bibfnamefont {C.}~\bibnamefont {Stra\ss{}er}},
  \bibinfo {author} {\bibfnamefont {K.}~\bibnamefont {Kern}}, \ and\ \bibinfo
  {author} {\bibfnamefont {C.~R.}\ \bibnamefont {Ast}},\ }\href@noop {}
  {\bibfield  {journal} {\bibinfo  {journal} {Phys. Rev. B}\ }\textbf {\bibinfo
  {volume} {91}},\ \bibinfo {pages} {161406} (\bibinfo {year}
  {2015})}\BibitemShut {NoStop}%
\bibitem [{\citenamefont {Qu}\ \emph {et~al.}(2010)\citenamefont {Qu},
  \citenamefont {Hor}, \citenamefont {Xiong}, \citenamefont {Cava},\ and\
  \citenamefont {Ong}}]{Qu821}%
  \BibitemOpen
  \bibfield  {author} {\bibinfo {author} {\bibfnamefont {D.-X.}\ \bibnamefont
  {Qu}}, \bibinfo {author} {\bibfnamefont {Y.~S.}\ \bibnamefont {Hor}},
  \bibinfo {author} {\bibfnamefont {J.}~\bibnamefont {Xiong}}, \bibinfo
  {author} {\bibfnamefont {R.~J.}\ \bibnamefont {Cava}}, \ and\ \bibinfo
  {author} {\bibfnamefont {N.~P.}\ \bibnamefont {Ong}},\ }\href@noop {}
  {\bibfield  {journal} {\bibinfo  {journal} {Science}\ }\textbf {\bibinfo
  {volume} {329}},\ \bibinfo {pages} {821} (\bibinfo {year}
  {2010})}\BibitemShut {NoStop}%
\bibitem [{\citenamefont {Analytis}\ \emph {et~al.}(2010)\citenamefont
  {Analytis}, \citenamefont {McDonald}, \citenamefont {Riggs}, \citenamefont
  {Chu}, \citenamefont {Boebinger},\ and\ \citenamefont
  {Fisher}}]{Analytis2010}%
  \BibitemOpen
  \bibfield  {author} {\bibinfo {author} {\bibfnamefont {J.~G.}\ \bibnamefont
  {Analytis}}, \bibinfo {author} {\bibfnamefont {R.~D.}\ \bibnamefont
  {McDonald}}, \bibinfo {author} {\bibfnamefont {S.~C.}\ \bibnamefont {Riggs}},
  \bibinfo {author} {\bibfnamefont {J.-H.}\ \bibnamefont {Chu}}, \bibinfo
  {author} {\bibfnamefont {G.~S.}\ \bibnamefont {Boebinger}}, \ and\ \bibinfo
  {author} {\bibfnamefont {I.~R.}\ \bibnamefont {Fisher}},\ }\href@noop {}
  {\bibfield  {journal} {\bibinfo  {journal} {Nat. Phys.}\ }\textbf {\bibinfo
  {volume} {6}},\ \bibinfo {pages} {960} (\bibinfo {year} {2010})}\BibitemShut
  {NoStop}%
\bibitem [{\citenamefont {Peng}\ \emph {et~al.}(2010)\citenamefont {Peng},
  \citenamefont {Lai}, \citenamefont {Kong}, \citenamefont {Meister},
  \citenamefont {Chen}, \citenamefont {Qi}, \citenamefont {Zhang},
  \citenamefont {Shen},\ and\ \citenamefont {Cui}}]{Peng2010}%
  \BibitemOpen
  \bibfield  {author} {\bibinfo {author} {\bibfnamefont {H.}~\bibnamefont
  {Peng}}, \bibinfo {author} {\bibfnamefont {K.}~\bibnamefont {Lai}}, \bibinfo
  {author} {\bibfnamefont {D.}~\bibnamefont {Kong}}, \bibinfo {author}
  {\bibfnamefont {S.}~\bibnamefont {Meister}}, \bibinfo {author} {\bibfnamefont
  {Y.}~\bibnamefont {Chen}}, \bibinfo {author} {\bibfnamefont {X.-L.}\
  \bibnamefont {Qi}}, \bibinfo {author} {\bibfnamefont {S.-C.}\ \bibnamefont
  {Zhang}}, \bibinfo {author} {\bibfnamefont {Z.-X.}\ \bibnamefont {Shen}}, \
  and\ \bibinfo {author} {\bibfnamefont {Y.}~\bibnamefont {Cui}},\ }\href@noop
  {} {\bibfield  {journal} {\bibinfo  {journal} {Nat. Mater.}\ }\textbf
  {\bibinfo {volume} {9}},\ \bibinfo {pages} {225} (\bibinfo {year}
  {2010})}\BibitemShut {NoStop}%
\bibitem [{\citenamefont {Chen}\ \emph {et~al.}(2010)\citenamefont {Chen},
  \citenamefont {Qin}, \citenamefont {Yang}, \citenamefont {Liu}, \citenamefont
  {Guan}, \citenamefont {Qu}, \citenamefont {Zhang}, \citenamefont {Shi},
  \citenamefont {Xie}, \citenamefont {Yang}, \citenamefont {Wu}, \citenamefont
  {Li},\ and\ \citenamefont {Lu}}]{PhysRevLett.105.176602}%
  \BibitemOpen
  \bibfield  {author} {\bibinfo {author} {\bibfnamefont {J.}~\bibnamefont
  {Chen}}, \bibinfo {author} {\bibfnamefont {H.~J.}\ \bibnamefont {Qin}},
  \bibinfo {author} {\bibfnamefont {F.}~\bibnamefont {Yang}}, \bibinfo {author}
  {\bibfnamefont {J.}~\bibnamefont {Liu}}, \bibinfo {author} {\bibfnamefont
  {T.}~\bibnamefont {Guan}}, \bibinfo {author} {\bibfnamefont {F.~M.}\
  \bibnamefont {Qu}}, \bibinfo {author} {\bibfnamefont {G.~H.}\ \bibnamefont
  {Zhang}}, \bibinfo {author} {\bibfnamefont {J.~R.}\ \bibnamefont {Shi}},
  \bibinfo {author} {\bibfnamefont {X.~C.}\ \bibnamefont {Xie}}, \bibinfo
  {author} {\bibfnamefont {C.~L.}\ \bibnamefont {Yang}}, \bibinfo {author}
  {\bibfnamefont {K.~H.}\ \bibnamefont {Wu}}, \bibinfo {author} {\bibfnamefont
  {Y.~Q.}\ \bibnamefont {Li}}, \ and\ \bibinfo {author} {\bibfnamefont
  {L.}~\bibnamefont {Lu}},\ }\href@noop {} {\bibfield  {journal} {\bibinfo
  {journal} {Phys. Rev. Lett.}\ }\textbf {\bibinfo {volume} {105}},\ \bibinfo
  {pages} {176602} (\bibinfo {year} {2010})}\BibitemShut {NoStop}%
\bibitem [{\citenamefont {Roushan}\ \emph {et~al.}(2009)\citenamefont
  {Roushan}, \citenamefont {Seo}, \citenamefont {Parker}, \citenamefont {Hor},
  \citenamefont {Hsieh}, \citenamefont {Qian}, \citenamefont {Richardella},
  \citenamefont {Hasan}, \citenamefont {Cava},\ and\ \citenamefont
  {Yazdani}}]{Roushan2009}%
  \BibitemOpen
  \bibfield  {author} {\bibinfo {author} {\bibfnamefont {P.}~\bibnamefont
  {Roushan}}, \bibinfo {author} {\bibfnamefont {J.}~\bibnamefont {Seo}},
  \bibinfo {author} {\bibfnamefont {C.~V.}\ \bibnamefont {Parker}}, \bibinfo
  {author} {\bibfnamefont {Y.~S.}\ \bibnamefont {Hor}}, \bibinfo {author}
  {\bibfnamefont {D.}~\bibnamefont {Hsieh}}, \bibinfo {author} {\bibfnamefont
  {D.}~\bibnamefont {Qian}}, \bibinfo {author} {\bibfnamefont {A.}~\bibnamefont
  {Richardella}}, \bibinfo {author} {\bibfnamefont {M.~Z.}\ \bibnamefont
  {Hasan}}, \bibinfo {author} {\bibfnamefont {R.~J.}\ \bibnamefont {Cava}}, \
  and\ \bibinfo {author} {\bibfnamefont {A.}~\bibnamefont {Yazdani}},\
  }\href@noop {} {\bibfield  {journal} {\bibinfo  {journal} {Nature}\ }\textbf
  {\bibinfo {volume} {460}},\ \bibinfo {pages} {1106} (\bibinfo {year}
  {2009})}\BibitemShut {NoStop}%
\bibitem [{\citenamefont {Zhang}\ \emph {et~al.}(2013)\citenamefont {Zhang},
  \citenamefont {Levy}, \citenamefont {Ha}, \citenamefont {Kuk},\ and\
  \citenamefont {Stroscio}}]{PhysRevB.87.115410}%
  \BibitemOpen
  \bibfield  {author} {\bibinfo {author} {\bibfnamefont {T.}~\bibnamefont
  {Zhang}}, \bibinfo {author} {\bibfnamefont {N.}~\bibnamefont {Levy}},
  \bibinfo {author} {\bibfnamefont {J.}~\bibnamefont {Ha}}, \bibinfo {author}
  {\bibfnamefont {Y.}~\bibnamefont {Kuk}}, \ and\ \bibinfo {author}
  {\bibfnamefont {J.~A.}\ \bibnamefont {Stroscio}},\ }\href@noop {} {\bibfield
  {journal} {\bibinfo  {journal} {Phys. Rev. B}\ }\textbf {\bibinfo {volume}
  {87}},\ \bibinfo {pages} {115410} (\bibinfo {year} {2013})}\BibitemShut
  {NoStop}%
\bibitem [{\citenamefont {Geim}\ and\ \citenamefont
  {Novoselov}(2007)}]{Geim2007}%
  \BibitemOpen
  \bibfield  {author} {\bibinfo {author} {\bibfnamefont {A.~K.}\ \bibnamefont
  {Geim}}\ and\ \bibinfo {author} {\bibfnamefont {K.~S.}\ \bibnamefont
  {Novoselov}},\ }\href@noop {} {\bibfield  {journal} {\bibinfo  {journal}
  {Nat. Mater.}\ }\textbf {\bibinfo {volume} {6}},\ \bibinfo {pages} {183}
  (\bibinfo {year} {2007})}\BibitemShut {NoStop}%
\bibitem [{\citenamefont {Durand}\ \emph {et~al.}(2016)\citenamefont {Durand},
  \citenamefont {Zhang}, \citenamefont {Hus}, \citenamefont {Ma}, \citenamefont
  {McGuire}, \citenamefont {Xu}, \citenamefont {Cao}, \citenamefont
  {Miotkowski}, \citenamefont {Chen},\ and\ \citenamefont
  {Li}}]{acs.nanolett.5b04425}%
  \BibitemOpen
  \bibfield  {author} {\bibinfo {author} {\bibfnamefont {C.}~\bibnamefont
  {Durand}}, \bibinfo {author} {\bibfnamefont {X.-G.}\ \bibnamefont {Zhang}},
  \bibinfo {author} {\bibfnamefont {S.~M.}\ \bibnamefont {Hus}}, \bibinfo
  {author} {\bibfnamefont {C.}~\bibnamefont {Ma}}, \bibinfo {author}
  {\bibfnamefont {M.~A.}\ \bibnamefont {McGuire}}, \bibinfo {author}
  {\bibfnamefont {Y.}~\bibnamefont {Xu}}, \bibinfo {author} {\bibfnamefont
  {H.}~\bibnamefont {Cao}}, \bibinfo {author} {\bibfnamefont {I.}~\bibnamefont
  {Miotkowski}}, \bibinfo {author} {\bibfnamefont {Y.~P.}\ \bibnamefont
  {Chen}}, \ and\ \bibinfo {author} {\bibfnamefont {A.-P.}\ \bibnamefont
  {Li}},\ }\href@noop {} {\bibfield  {journal} {\bibinfo  {journal} {Nano
  Lett.}\ }\textbf {\bibinfo {volume} {16}},\ \bibinfo {pages} {2213} (\bibinfo
  {year} {2016})}\BibitemShut {NoStop}%
\bibitem [{\citenamefont {Jia}\ \emph {et~al.}(2011)\citenamefont {Jia},
  \citenamefont {Ji}, \citenamefont {Climent-Pascual}, \citenamefont
  {Fuccillo}, \citenamefont {Charles}, \citenamefont {Xiong}, \citenamefont
  {Ong},\ and\ \citenamefont {Cava}}]{PhysRevB.84.235206}%
  \BibitemOpen
  \bibfield  {author} {\bibinfo {author} {\bibfnamefont {S.}~\bibnamefont
  {Jia}}, \bibinfo {author} {\bibfnamefont {H.}~\bibnamefont {Ji}}, \bibinfo
  {author} {\bibfnamefont {E.}~\bibnamefont {Climent-Pascual}}, \bibinfo
  {author} {\bibfnamefont {M.~K.}\ \bibnamefont {Fuccillo}}, \bibinfo {author}
  {\bibfnamefont {M.~E.}\ \bibnamefont {Charles}}, \bibinfo {author}
  {\bibfnamefont {J.}~\bibnamefont {Xiong}}, \bibinfo {author} {\bibfnamefont
  {N.~P.}\ \bibnamefont {Ong}}, \ and\ \bibinfo {author} {\bibfnamefont
  {R.~J.}\ \bibnamefont {Cava}},\ }\href@noop {} {\bibfield  {journal}
  {\bibinfo  {journal} {Phys. Rev. B}\ }\textbf {\bibinfo {volume} {84}},\
  \bibinfo {pages} {235206} (\bibinfo {year} {2011})}\BibitemShut {NoStop}%
\bibitem [{\citenamefont {Nurmamat}\ \emph {et~al.}(2013)\citenamefont
  {Nurmamat}, \citenamefont {Krasovskii}, \citenamefont {Kuroda}, \citenamefont
  {Ye}, \citenamefont {Miyamoto}, \citenamefont {Nakatake}, \citenamefont
  {Okuda}, \citenamefont {Namatame}, \citenamefont {Taniguchi}, \citenamefont
  {Chulkov}, \citenamefont {Kokh}, \citenamefont {Tereshchenko},\ and\
  \citenamefont {Kimura}}]{PhysRevB.88.081301}%
  \BibitemOpen
  \bibfield  {author} {\bibinfo {author} {\bibfnamefont {M.}~\bibnamefont
  {Nurmamat}}, \bibinfo {author} {\bibfnamefont {E.~E.}\ \bibnamefont
  {Krasovskii}}, \bibinfo {author} {\bibfnamefont {K.}~\bibnamefont {Kuroda}},
  \bibinfo {author} {\bibfnamefont {M.}~\bibnamefont {Ye}}, \bibinfo {author}
  {\bibfnamefont {K.}~\bibnamefont {Miyamoto}}, \bibinfo {author}
  {\bibfnamefont {M.}~\bibnamefont {Nakatake}}, \bibinfo {author}
  {\bibfnamefont {T.}~\bibnamefont {Okuda}}, \bibinfo {author} {\bibfnamefont
  {H.}~\bibnamefont {Namatame}}, \bibinfo {author} {\bibfnamefont
  {M.}~\bibnamefont {Taniguchi}}, \bibinfo {author} {\bibfnamefont {E.~V.}\
  \bibnamefont {Chulkov}}, \bibinfo {author} {\bibfnamefont {K.~A.}\
  \bibnamefont {Kokh}}, \bibinfo {author} {\bibfnamefont {O.~E.}\ \bibnamefont
  {Tereshchenko}}, \ and\ \bibinfo {author} {\bibfnamefont {A.}~\bibnamefont
  {Kimura}},\ }\href@noop {} {\bibfield  {journal} {\bibinfo  {journal} {Phys.
  Rev. B}\ }\textbf {\bibinfo {volume} {88}},\ \bibinfo {pages} {081301}
  (\bibinfo {year} {2013})}\BibitemShut {NoStop}%
\bibitem [{\citenamefont {Shekhar}\ \emph {et~al.}(2014)\citenamefont
  {Shekhar}, \citenamefont {ViolBarbosa}, \citenamefont {Yan}, \citenamefont
  {Ouardi}, \citenamefont {Schnelle}, \citenamefont {Fecher},\ and\
  \citenamefont {Felser}}]{PhysRevB.90.165140}%
  \BibitemOpen
  \bibfield  {author} {\bibinfo {author} {\bibfnamefont {C.}~\bibnamefont
  {Shekhar}}, \bibinfo {author} {\bibfnamefont {C.~E.}\ \bibnamefont
  {ViolBarbosa}}, \bibinfo {author} {\bibfnamefont {B.}~\bibnamefont {Yan}},
  \bibinfo {author} {\bibfnamefont {S.}~\bibnamefont {Ouardi}}, \bibinfo
  {author} {\bibfnamefont {W.}~\bibnamefont {Schnelle}}, \bibinfo {author}
  {\bibfnamefont {G.~H.}\ \bibnamefont {Fecher}}, \ and\ \bibinfo {author}
  {\bibfnamefont {C.}~\bibnamefont {Felser}},\ }\href@noop {} {\bibfield
  {journal} {\bibinfo  {journal} {Phys. Rev. B}\ }\textbf {\bibinfo {volume}
  {90}},\ \bibinfo {pages} {165140} (\bibinfo {year} {2014})}\BibitemShut
  {NoStop}%
\bibitem [{\citenamefont {Shikin}\ \emph {et~al.}(2014)\citenamefont {Shikin},
  \citenamefont {Klimovskikh}, \citenamefont {Eremeev}, \citenamefont
  {Rybkina}, \citenamefont {Rusinova}, \citenamefont {Rybkin}, \citenamefont
  {Zhizhin}, \citenamefont {S\'anchez-Barriga}, \citenamefont {Varykhalov},
  \citenamefont {Rusinov}, \citenamefont {Chulkov}, \citenamefont {Kokh},
  \citenamefont {Golyashov}, \citenamefont {Kamyshlov},\ and\ \citenamefont
  {Tereshchenko}}]{PhysRevB.89.125416}%
  \BibitemOpen
  \bibfield  {author} {\bibinfo {author} {\bibfnamefont {A.~M.}\ \bibnamefont
  {Shikin}}, \bibinfo {author} {\bibfnamefont {I.~I.}\ \bibnamefont
  {Klimovskikh}}, \bibinfo {author} {\bibfnamefont {S.~V.}\ \bibnamefont
  {Eremeev}}, \bibinfo {author} {\bibfnamefont {A.~A.}\ \bibnamefont
  {Rybkina}}, \bibinfo {author} {\bibfnamefont {M.~V.}\ \bibnamefont
  {Rusinova}}, \bibinfo {author} {\bibfnamefont {A.~G.}\ \bibnamefont
  {Rybkin}}, \bibinfo {author} {\bibfnamefont {E.~V.}\ \bibnamefont {Zhizhin}},
  \bibinfo {author} {\bibfnamefont {J.}~\bibnamefont {S\'anchez-Barriga}},
  \bibinfo {author} {\bibfnamefont {A.}~\bibnamefont {Varykhalov}}, \bibinfo
  {author} {\bibfnamefont {I.~P.}\ \bibnamefont {Rusinov}}, \bibinfo {author}
  {\bibfnamefont {E.~V.}\ \bibnamefont {Chulkov}}, \bibinfo {author}
  {\bibfnamefont {K.~A.}\ \bibnamefont {Kokh}}, \bibinfo {author}
  {\bibfnamefont {V.~A.}\ \bibnamefont {Golyashov}}, \bibinfo {author}
  {\bibfnamefont {V.}~\bibnamefont {Kamyshlov}}, \ and\ \bibinfo {author}
  {\bibfnamefont {O.~E.}\ \bibnamefont {Tereshchenko}},\ }\href@noop {}
  {\bibfield  {journal} {\bibinfo  {journal} {Phys. Rev. B}\ }\textbf {\bibinfo
  {volume} {89}},\ \bibinfo {pages} {125416} (\bibinfo {year}
  {2014})}\BibitemShut {NoStop}%
\bibitem [{\citenamefont {Zhang}\ \emph {et~al.}(2011)\citenamefont {Zhang},
  \citenamefont {Chang}, \citenamefont {Zhang}, \citenamefont {Wen},
  \citenamefont {Feng}, \citenamefont {Li}, \citenamefont {Liu}, \citenamefont
  {He}, \citenamefont {Wang}, \citenamefont {Chen}, \citenamefont {Xue},
  \citenamefont {Ma},\ and\ \citenamefont {Wang}}]{Zhang2011}%
  \BibitemOpen
  \bibfield  {author} {\bibinfo {author} {\bibfnamefont {J.}~\bibnamefont
  {Zhang}}, \bibinfo {author} {\bibfnamefont {C.-Z.}\ \bibnamefont {Chang}},
  \bibinfo {author} {\bibfnamefont {Z.}~\bibnamefont {Zhang}}, \bibinfo
  {author} {\bibfnamefont {J.}~\bibnamefont {Wen}}, \bibinfo {author}
  {\bibfnamefont {X.}~\bibnamefont {Feng}}, \bibinfo {author} {\bibfnamefont
  {K.}~\bibnamefont {Li}}, \bibinfo {author} {\bibfnamefont {M.}~\bibnamefont
  {Liu}}, \bibinfo {author} {\bibfnamefont {K.}~\bibnamefont {He}}, \bibinfo
  {author} {\bibfnamefont {L.}~\bibnamefont {Wang}}, \bibinfo {author}
  {\bibfnamefont {X.}~\bibnamefont {Chen}}, \bibinfo {author} {\bibfnamefont
  {Q.-K.}\ \bibnamefont {Xue}}, \bibinfo {author} {\bibfnamefont
  {X.}~\bibnamefont {Ma}}, \ and\ \bibinfo {author} {\bibfnamefont
  {Y.}~\bibnamefont {Wang}},\ }\href@noop {} {\bibfield  {journal} {\bibinfo
  {journal} {Nat. Commun.}\ }\textbf {\bibinfo {volume} {2}},\ \bibinfo {pages}
  {574} (\bibinfo {year} {2011})}\BibitemShut {NoStop}%
\bibitem [{\citenamefont {Kong}\ \emph {et~al.}(2011)\citenamefont {Kong},
  \citenamefont {Chen}, \citenamefont {Cha}, \citenamefont {Zhang},
  \citenamefont {Analytis}, \citenamefont {Lai}, \citenamefont {Liu},
  \citenamefont {Hong}, \citenamefont {Koski}, \citenamefont {Mo},
  \citenamefont {Hussain}, \citenamefont {Fisher}, \citenamefont {Shen},\ and\
  \citenamefont {Cui}}]{Kong2011}%
  \BibitemOpen
  \bibfield  {author} {\bibinfo {author} {\bibfnamefont {D.}~\bibnamefont
  {Kong}}, \bibinfo {author} {\bibfnamefont {Y.}~\bibnamefont {Chen}}, \bibinfo
  {author} {\bibfnamefont {J.~J.}\ \bibnamefont {Cha}}, \bibinfo {author}
  {\bibfnamefont {Q.}~\bibnamefont {Zhang}}, \bibinfo {author} {\bibfnamefont
  {J.~G.}\ \bibnamefont {Analytis}}, \bibinfo {author} {\bibfnamefont
  {K.}~\bibnamefont {Lai}}, \bibinfo {author} {\bibfnamefont {Z.}~\bibnamefont
  {Liu}}, \bibinfo {author} {\bibfnamefont {S.~S.}\ \bibnamefont {Hong}},
  \bibinfo {author} {\bibfnamefont {K.~J.}\ \bibnamefont {Koski}}, \bibinfo
  {author} {\bibfnamefont {S.-K.}\ \bibnamefont {Mo}}, \bibinfo {author}
  {\bibfnamefont {Z.}~\bibnamefont {Hussain}}, \bibinfo {author} {\bibfnamefont
  {I.~R.}\ \bibnamefont {Fisher}}, \bibinfo {author} {\bibfnamefont {Z.-X.}\
  \bibnamefont {Shen}}, \ and\ \bibinfo {author} {\bibfnamefont
  {Y.}~\bibnamefont {Cui}},\ }\href@noop {} {\bibfield  {journal} {\bibinfo
  {journal} {Nat. Nanotechnol.}\ }\textbf {\bibinfo {volume} {6}},\ \bibinfo
  {pages} {705} (\bibinfo {year} {2011})}\BibitemShut {NoStop}%
\bibitem [{\citenamefont {Weyrich}\ \emph {et~al.}(2016)\citenamefont
  {Weyrich}, \citenamefont {Drögeler}, \citenamefont {Kampmeier},
  \citenamefont {Eschbach}, \citenamefont {Mussler}, \citenamefont {Merzenich},
  \citenamefont {Stoica}, \citenamefont {Batov}, \citenamefont {Schubert},
  \citenamefont {Plucinski}, \citenamefont {Beschoten}, \citenamefont
  {Schneider}, \citenamefont {Stampfer}, \citenamefont {Grützmacher},\ and\
  \citenamefont {Schäpers}}]{0953-8984-28-49-495501}%
  \BibitemOpen
  \bibfield  {author} {\bibinfo {author} {\bibfnamefont {C.}~\bibnamefont
  {Weyrich}}, \bibinfo {author} {\bibfnamefont {M.}~\bibnamefont {Drögeler}},
  \bibinfo {author} {\bibfnamefont {J.}~\bibnamefont {Kampmeier}}, \bibinfo
  {author} {\bibfnamefont {M.}~\bibnamefont {Eschbach}}, \bibinfo {author}
  {\bibfnamefont {G.}~\bibnamefont {Mussler}}, \bibinfo {author} {\bibfnamefont
  {T.}~\bibnamefont {Merzenich}}, \bibinfo {author} {\bibfnamefont
  {T.}~\bibnamefont {Stoica}}, \bibinfo {author} {\bibfnamefont {I.~E.}\
  \bibnamefont {Batov}}, \bibinfo {author} {\bibfnamefont {J.}~\bibnamefont
  {Schubert}}, \bibinfo {author} {\bibfnamefont {L.}~\bibnamefont {Plucinski}},
  \bibinfo {author} {\bibfnamefont {B.}~\bibnamefont {Beschoten}}, \bibinfo
  {author} {\bibfnamefont {C.~M.}\ \bibnamefont {Schneider}}, \bibinfo {author}
  {\bibfnamefont {C.}~\bibnamefont {Stampfer}}, \bibinfo {author}
  {\bibfnamefont {D.}~\bibnamefont {Grützmacher}}, \ and\ \bibinfo {author}
  {\bibfnamefont {T.}~\bibnamefont {Schäpers}},\ }\href@noop {} {\bibfield
  {journal} {\bibinfo  {journal} {J. Phys. Condens. Matter}\ }\textbf {\bibinfo
  {volume} {28}},\ \bibinfo {pages} {495501} (\bibinfo {year}
  {2016})}\BibitemShut {NoStop}%
\bibitem [{\citenamefont {Arakane}\ \emph {et~al.}(2012)\citenamefont
  {Arakane}, \citenamefont {Sato}, \citenamefont {Souma}, \citenamefont
  {Kosaka}, \citenamefont {Nakayama}, \citenamefont {Komatsu}, \citenamefont
  {Takahashi}, \citenamefont {Ren}, \citenamefont {Segawa},\ and\ \citenamefont
  {Ando}}]{Arakane2012}%
  \BibitemOpen
  \bibfield  {author} {\bibinfo {author} {\bibfnamefont {T.}~\bibnamefont
  {Arakane}}, \bibinfo {author} {\bibfnamefont {T.}~\bibnamefont {Sato}},
  \bibinfo {author} {\bibfnamefont {S.}~\bibnamefont {Souma}}, \bibinfo
  {author} {\bibfnamefont {K.}~\bibnamefont {Kosaka}}, \bibinfo {author}
  {\bibfnamefont {K.}~\bibnamefont {Nakayama}}, \bibinfo {author}
  {\bibfnamefont {M.}~\bibnamefont {Komatsu}}, \bibinfo {author} {\bibfnamefont
  {T.}~\bibnamefont {Takahashi}}, \bibinfo {author} {\bibfnamefont
  {Z.}~\bibnamefont {Ren}}, \bibinfo {author} {\bibfnamefont {K.}~\bibnamefont
  {Segawa}}, \ and\ \bibinfo {author} {\bibfnamefont {Y.}~\bibnamefont
  {Ando}},\ }\href@noop {} {\bibfield  {journal} {\bibinfo  {journal} {Nat.
  Commun.}\ }\textbf {\bibinfo {volume} {3}},\ \bibinfo {pages} {636} (\bibinfo
  {year} {2012})}\BibitemShut {NoStop}%
\bibitem [{\citenamefont {Kushwaha}\ \emph {et~al.}(2016)\citenamefont
  {Kushwaha}, \citenamefont {Pletikosic}, \citenamefont {Liang}, \citenamefont
  {Gyenis}, \citenamefont {Lapidus}, \citenamefont {Tian}, \citenamefont
  {Zhao}, \citenamefont {Burch}, \citenamefont {Lin}, \citenamefont {Wang},
  \citenamefont {Ji}, \citenamefont {Fedorov}, \citenamefont {Yazdani},
  \citenamefont {Ong}, \citenamefont {Valla},\ and\ \citenamefont
  {Cava}}]{Kushwaha2016}%
  \BibitemOpen
  \bibfield  {author} {\bibinfo {author} {\bibfnamefont {S.~K.}\ \bibnamefont
  {Kushwaha}}, \bibinfo {author} {\bibfnamefont {I.}~\bibnamefont
  {Pletikosic}}, \bibinfo {author} {\bibfnamefont {T.}~\bibnamefont {Liang}},
  \bibinfo {author} {\bibfnamefont {A.}~\bibnamefont {Gyenis}}, \bibinfo
  {author} {\bibfnamefont {S.~H.}\ \bibnamefont {Lapidus}}, \bibinfo {author}
  {\bibfnamefont {Y.}~\bibnamefont {Tian}}, \bibinfo {author} {\bibfnamefont
  {H.}~\bibnamefont {Zhao}}, \bibinfo {author} {\bibfnamefont {K.~S.}\
  \bibnamefont {Burch}}, \bibinfo {author} {\bibfnamefont {J.}~\bibnamefont
  {Lin}}, \bibinfo {author} {\bibfnamefont {W.}~\bibnamefont {Wang}}, \bibinfo
  {author} {\bibfnamefont {H.}~\bibnamefont {Ji}}, \bibinfo {author}
  {\bibfnamefont {A.~V.}\ \bibnamefont {Fedorov}}, \bibinfo {author}
  {\bibfnamefont {A.}~\bibnamefont {Yazdani}}, \bibinfo {author} {\bibfnamefont
  {N.~P.}\ \bibnamefont {Ong}}, \bibinfo {author} {\bibfnamefont
  {T.}~\bibnamefont {Valla}}, \ and\ \bibinfo {author} {\bibfnamefont {R.~J.}\
  \bibnamefont {Cava}},\ }\href@noop {} {\bibfield  {journal} {\bibinfo
  {journal} {Nat. Commun.}\ }\textbf {\bibinfo {volume} {7}},\ \bibinfo {pages}
  {11456} (\bibinfo {year} {2016})}\BibitemShut {NoStop}%
\bibitem [{\citenamefont {Fu}\ and\ \citenamefont
  {Kane}(2007)}]{PhysRevB.76.045302}%
  \BibitemOpen
  \bibfield  {author} {\bibinfo {author} {\bibfnamefont {L.}~\bibnamefont
  {Fu}}\ and\ \bibinfo {author} {\bibfnamefont {C.~L.}\ \bibnamefont {Kane}},\
  }\href@noop {} {\bibfield  {journal} {\bibinfo  {journal} {Phys. Rev. B}\
  }\textbf {\bibinfo {volume} {76}},\ \bibinfo {pages} {045302} (\bibinfo
  {year} {2007})}\BibitemShut {NoStop}%
\bibitem [{\citenamefont {Bian}\ \emph {et~al.}(2011)\citenamefont {Bian},
  \citenamefont {Miller},\ and\ \citenamefont
  {Chiang}}]{PhysRevLett.107.036802}%
  \BibitemOpen
  \bibfield  {author} {\bibinfo {author} {\bibfnamefont {G.}~\bibnamefont
  {Bian}}, \bibinfo {author} {\bibfnamefont {T.}~\bibnamefont {Miller}}, \ and\
  \bibinfo {author} {\bibfnamefont {T.-C.}\ \bibnamefont {Chiang}},\
  }\href@noop {} {\bibfield  {journal} {\bibinfo  {journal} {Phys. Rev. Lett.}\
  }\textbf {\bibinfo {volume} {107}},\ \bibinfo {pages} {036802} (\bibinfo
  {year} {2011})}\BibitemShut {NoStop}%
\bibitem [{\citenamefont {Zhang}\ \emph {et~al.}(2015)\citenamefont {Zhang},
  \citenamefont {Yan}, \citenamefont {Li}, \citenamefont {Chen},\ and\
  \citenamefont {Zeng}}]{ANIE:ANIE201411246}%
  \BibitemOpen
  \bibfield  {author} {\bibinfo {author} {\bibfnamefont {S.}~\bibnamefont
  {Zhang}}, \bibinfo {author} {\bibfnamefont {Z.}~\bibnamefont {Yan}}, \bibinfo
  {author} {\bibfnamefont {Y.}~\bibnamefont {Li}}, \bibinfo {author}
  {\bibfnamefont {Z.}~\bibnamefont {Chen}}, \ and\ \bibinfo {author}
  {\bibfnamefont {H.}~\bibnamefont {Zeng}},\ }\href@noop {} {\bibfield
  {journal} {\bibinfo  {journal} {Angew. Chem. Int. Ed.}\ }\textbf {\bibinfo
  {volume} {54}},\ \bibinfo {pages} {3112} (\bibinfo {year}
  {2015})}\BibitemShut {NoStop}%
\bibitem [{\citenamefont {Ares}\ \emph {et~al.}(2016)\citenamefont {Ares},
  \citenamefont {Aguilar-Galindo}, \citenamefont {Rodr\'iguez-San-Miguel},
  \citenamefont {Aldave}, \citenamefont {D\'iaz-Tendero}, \citenamefont
  {Alcam\'i}, \citenamefont {Mart\'in}, \citenamefont {G\'omez-Herrero},\ and\
  \citenamefont {Zamora}}]{ADMA:ADMA201602128}%
  \BibitemOpen
  \bibfield  {author} {\bibinfo {author} {\bibfnamefont {P.}~\bibnamefont
  {Ares}}, \bibinfo {author} {\bibfnamefont {F.}~\bibnamefont
  {Aguilar-Galindo}}, \bibinfo {author} {\bibfnamefont {D.}~\bibnamefont
  {Rodr\'iguez-San-Miguel}}, \bibinfo {author} {\bibfnamefont {D.~A.}\
  \bibnamefont {Aldave}}, \bibinfo {author} {\bibfnamefont {S.}~\bibnamefont
  {D\'iaz-Tendero}}, \bibinfo {author} {\bibfnamefont {M.}~\bibnamefont
  {Alcam\'i}}, \bibinfo {author} {\bibfnamefont {F.}~\bibnamefont {Mart\'in}},
  \bibinfo {author} {\bibfnamefont {J.}~\bibnamefont {G\'omez-Herrero}}, \ and\
  \bibinfo {author} {\bibfnamefont {F.}~\bibnamefont {Zamora}},\ }\href@noop {}
  {\bibfield  {journal} {\bibinfo  {journal} {Adv. Mater.}\ }\textbf {\bibinfo
  {volume} {28}},\ \bibinfo {pages} {6332} (\bibinfo {year}
  {2016})}\BibitemShut {NoStop}%
\bibitem [{\citenamefont {Castellanos-Gomez}\ \emph {et~al.}(2014)\citenamefont
  {Castellanos-Gomez}, \citenamefont {Vicarelli}, \citenamefont {Prada},
  \citenamefont {Island}, \citenamefont {Narasimha-Acharya}, \citenamefont
  {Blanter}, \citenamefont {Groenendijk}, \citenamefont {Buscema},
  \citenamefont {Steele}, \citenamefont {Alvarez}, \citenamefont {Zandbergen},
  \citenamefont {Palacios},\ and\ \citenamefont {van~der Zant}}]{phosphorene}%
  \BibitemOpen
  \bibfield  {author} {\bibinfo {author} {\bibfnamefont {A.}~\bibnamefont
  {Castellanos-Gomez}}, \bibinfo {author} {\bibfnamefont {L.}~\bibnamefont
  {Vicarelli}}, \bibinfo {author} {\bibfnamefont {E.}~\bibnamefont {Prada}},
  \bibinfo {author} {\bibfnamefont {J.~O.}\ \bibnamefont {Island}}, \bibinfo
  {author} {\bibfnamefont {K.~L.}\ \bibnamefont {Narasimha-Acharya}}, \bibinfo
  {author} {\bibfnamefont {S.~I.}\ \bibnamefont {Blanter}}, \bibinfo {author}
  {\bibfnamefont {D.~J.}\ \bibnamefont {Groenendijk}}, \bibinfo {author}
  {\bibfnamefont {M.}~\bibnamefont {Buscema}}, \bibinfo {author} {\bibfnamefont
  {G.~A.}\ \bibnamefont {Steele}}, \bibinfo {author} {\bibfnamefont {J.~V.}\
  \bibnamefont {Alvarez}}, \bibinfo {author} {\bibfnamefont {H.~W.}\
  \bibnamefont {Zandbergen}}, \bibinfo {author} {\bibfnamefont {J.~J.}\
  \bibnamefont {Palacios}}, \ and\ \bibinfo {author} {\bibfnamefont {H.~S.~J.}\
  \bibnamefont {van~der Zant}},\ }\href@noop {} {\bibfield  {journal} {\bibinfo
   {journal} {2D Mater.}\ }\textbf {\bibinfo {volume} {1}},\ \bibinfo {pages}
  {025001} (\bibinfo {year} {2014})}\BibitemShut {NoStop}%
\bibitem [{\citenamefont {Zhang}\ \emph {et~al.}(2012)\citenamefont {Zhang},
  \citenamefont {Liu}, \citenamefont {Duan}, \citenamefont {Liu},\ and\
  \citenamefont {Wu}}]{PhysRevB.85.201410}%
  \BibitemOpen
  \bibfield  {author} {\bibinfo {author} {\bibfnamefont {P.}~\bibnamefont
  {Zhang}}, \bibinfo {author} {\bibfnamefont {Z.}~\bibnamefont {Liu}}, \bibinfo
  {author} {\bibfnamefont {W.}~\bibnamefont {Duan}}, \bibinfo {author}
  {\bibfnamefont {F.}~\bibnamefont {Liu}}, \ and\ \bibinfo {author}
  {\bibfnamefont {J.}~\bibnamefont {Wu}},\ }\href@noop {} {\bibfield  {journal}
  {\bibinfo  {journal} {Phys. Rev. B}\ }\textbf {\bibinfo {volume} {85}},\
  \bibinfo {pages} {201410} (\bibinfo {year} {2012})}\BibitemShut {NoStop}%
\bibitem [{\citenamefont {Bian}\ \emph {et~al.}(2012)\citenamefont {Bian},
  \citenamefont {Wang}, \citenamefont {Liu}, \citenamefont {Miller},\ and\
  \citenamefont {Chiang}}]{PhysRevLett.108.176401}%
  \BibitemOpen
  \bibfield  {author} {\bibinfo {author} {\bibfnamefont {G.}~\bibnamefont
  {Bian}}, \bibinfo {author} {\bibfnamefont {X.}~\bibnamefont {Wang}}, \bibinfo
  {author} {\bibfnamefont {Y.}~\bibnamefont {Liu}}, \bibinfo {author}
  {\bibfnamefont {T.}~\bibnamefont {Miller}}, \ and\ \bibinfo {author}
  {\bibfnamefont {T.-C.}\ \bibnamefont {Chiang}},\ }\href@noop {} {\bibfield
  {journal} {\bibinfo  {journal} {Phys. Rev. Lett.}\ }\textbf {\bibinfo
  {volume} {108}},\ \bibinfo {pages} {176401} (\bibinfo {year}
  {2012})}\BibitemShut {NoStop}%
\bibitem [{\citenamefont {Markoš}\ and\ \citenamefont
  {Schweitzer}(2006)}]{0305-4470-39-13-003}%
  \BibitemOpen
  \bibfield  {author} {\bibinfo {author} {\bibfnamefont {P.}~\bibnamefont
  {Markoš}}\ and\ \bibinfo {author} {\bibfnamefont {L.}~\bibnamefont
  {Schweitzer}},\ }\href@noop {} {\bibfield  {journal} {\bibinfo  {journal} {J.
  Phys. A Math. Gen.}\ }\textbf {\bibinfo {volume} {39}},\ \bibinfo {pages}
  {3221} (\bibinfo {year} {2006})}\BibitemShut {NoStop}%
\bibitem [{\citenamefont {Asada}\ \emph {et~al.}(2002)\citenamefont {Asada},
  \citenamefont {Slevin},\ and\ \citenamefont
  {Ohtsuki}}]{PhysRevLett.89.256601}%
  \BibitemOpen
  \bibfield  {author} {\bibinfo {author} {\bibfnamefont {Y.}~\bibnamefont
  {Asada}}, \bibinfo {author} {\bibfnamefont {K.}~\bibnamefont {Slevin}}, \
  and\ \bibinfo {author} {\bibfnamefont {T.}~\bibnamefont {Ohtsuki}},\
  }\href@noop {} {\bibfield  {journal} {\bibinfo  {journal} {Phys. Rev. Lett.}\
  }\textbf {\bibinfo {volume} {89}},\ \bibinfo {pages} {256601} (\bibinfo
  {year} {2002})}\BibitemShut {NoStop}%
\bibitem [{\citenamefont {Abrahams}\ \emph {et~al.}(1979)\citenamefont
  {Abrahams}, \citenamefont {Anderson}, \citenamefont {Licciardello},\ and\
  \citenamefont {Ramakrishnan}}]{PhysRevLett.42.673}%
  \BibitemOpen
  \bibfield  {author} {\bibinfo {author} {\bibfnamefont {E.}~\bibnamefont
  {Abrahams}}, \bibinfo {author} {\bibfnamefont {P.~W.}\ \bibnamefont
  {Anderson}}, \bibinfo {author} {\bibfnamefont {D.~C.}\ \bibnamefont
  {Licciardello}}, \ and\ \bibinfo {author} {\bibfnamefont {T.~V.}\
  \bibnamefont {Ramakrishnan}},\ }\href@noop {} {\bibfield  {journal} {\bibinfo
   {journal} {Phys. Rev. Lett.}\ }\textbf {\bibinfo {volume} {42}},\ \bibinfo
  {pages} {673} (\bibinfo {year} {1979})}\BibitemShut {NoStop}%
\bibitem [{\citenamefont {Bardarson}\ \emph {et~al.}(2007)\citenamefont
  {Bardarson}, \citenamefont {Tworzyd\l{}o}, \citenamefont {Brouwer},\ and\
  \citenamefont {Beenakker}}]{PhysRevLett.99.106801}%
  \BibitemOpen
  \bibfield  {author} {\bibinfo {author} {\bibfnamefont {J.~H.}\ \bibnamefont
  {Bardarson}}, \bibinfo {author} {\bibfnamefont {J.}~\bibnamefont
  {Tworzyd\l{}o}}, \bibinfo {author} {\bibfnamefont {P.~W.}\ \bibnamefont
  {Brouwer}}, \ and\ \bibinfo {author} {\bibfnamefont {C.~W.~J.}\ \bibnamefont
  {Beenakker}},\ }\href@noop {} {\bibfield  {journal} {\bibinfo  {journal}
  {Phys. Rev. Lett.}\ }\textbf {\bibinfo {volume} {99}},\ \bibinfo {pages}
  {106801} (\bibinfo {year} {2007})}\BibitemShut {NoStop}%
\bibitem [{\citenamefont {Terletska}\ \emph {et~al.}(2018)\citenamefont
  {Terletska}, \citenamefont {Zhang}, \citenamefont {Tam}, \citenamefont
  {Berlijn}, \citenamefont {Chioncel}, \citenamefont {Vidhyadhiraja},\ and\
  \citenamefont {Jarrell}}]{terletska2018systematic}%
  \BibitemOpen
  \bibfield  {author} {\bibinfo {author} {\bibfnamefont {H.}~\bibnamefont
  {Terletska}}, \bibinfo {author} {\bibfnamefont {Y.}~\bibnamefont {Zhang}},
  \bibinfo {author} {\bibfnamefont {K.-M.}\ \bibnamefont {Tam}}, \bibinfo
  {author} {\bibfnamefont {T.}~\bibnamefont {Berlijn}}, \bibinfo {author}
  {\bibfnamefont {L.}~\bibnamefont {Chioncel}}, \bibinfo {author}
  {\bibfnamefont {N.}~\bibnamefont {Vidhyadhiraja}}, \ and\ \bibinfo {author}
  {\bibfnamefont {M.}~\bibnamefont {Jarrell}},\ }\href@noop {} {\bibfield
  {journal} {\bibinfo  {journal} {Appl. Sci.}\ }\textbf {\bibinfo {volume}
  {8}},\ \bibinfo {pages} {2401} (\bibinfo {year} {2018})}\BibitemShut
  {NoStop}%
\bibitem [{\citenamefont {Dovesi}\ \emph {et~al.}(2014)\citenamefont {Dovesi},
  \citenamefont {Orlando}, \citenamefont {Erba}, \citenamefont
  {Zicovich-Wilson}, \citenamefont {Civalleri}, \citenamefont {Casassa},
  \citenamefont {Maschio}, \citenamefont {Ferrabone}, \citenamefont
  {De~La~Pierre}, \citenamefont {D'Arco} \emph {et~al.}}]{dovesi2014crystal14}%
  \BibitemOpen
  \bibfield  {author} {\bibinfo {author} {\bibfnamefont {R.}~\bibnamefont
  {Dovesi}}, \bibinfo {author} {\bibfnamefont {R.}~\bibnamefont {Orlando}},
  \bibinfo {author} {\bibfnamefont {A.}~\bibnamefont {Erba}}, \bibinfo {author}
  {\bibfnamefont {C.~M.}\ \bibnamefont {Zicovich-Wilson}}, \bibinfo {author}
  {\bibfnamefont {B.}~\bibnamefont {Civalleri}}, \bibinfo {author}
  {\bibfnamefont {S.}~\bibnamefont {Casassa}}, \bibinfo {author} {\bibfnamefont
  {L.}~\bibnamefont {Maschio}}, \bibinfo {author} {\bibfnamefont
  {M.}~\bibnamefont {Ferrabone}}, \bibinfo {author} {\bibfnamefont
  {M.}~\bibnamefont {De~La~Pierre}}, \bibinfo {author} {\bibfnamefont
  {P.}~\bibnamefont {D'Arco}},  \emph {et~al.},\ }\href@noop {} {\bibfield
  {journal} {\bibinfo  {journal} {Int. J. Quantum Chem.}\ }\textbf {\bibinfo
  {volume} {114}},\ \bibinfo {pages} {1287} (\bibinfo {year}
  {2014})}\BibitemShut {NoStop}%
\bibitem [{\citenamefont {Dovesi}\ \emph {et~al.}()\citenamefont {Dovesi},
  \citenamefont {Saunders}, \citenamefont {Roetti}, \citenamefont {Orlando},
  \citenamefont {Zicovich-Wilson}, \citenamefont {Pascale}, \citenamefont
  {Civalleri}, \citenamefont {Doll}, \citenamefont {Harrison}, \citenamefont
  {Bush} \emph {et~al.}}]{dovesicrystal14}%
  \BibitemOpen
  \bibfield  {author} {\bibinfo {author} {\bibfnamefont {R.}~\bibnamefont
  {Dovesi}}, \bibinfo {author} {\bibfnamefont {V.}~\bibnamefont {Saunders}},
  \bibinfo {author} {\bibfnamefont {C.}~\bibnamefont {Roetti}}, \bibinfo
  {author} {\bibfnamefont {R.}~\bibnamefont {Orlando}}, \bibinfo {author}
  {\bibfnamefont {C.}~\bibnamefont {Zicovich-Wilson}}, \bibinfo {author}
  {\bibfnamefont {F.}~\bibnamefont {Pascale}}, \bibinfo {author} {\bibfnamefont
  {B.}~\bibnamefont {Civalleri}}, \bibinfo {author} {\bibfnamefont
  {K.}~\bibnamefont {Doll}}, \bibinfo {author} {\bibfnamefont {N.}~\bibnamefont
  {Harrison}}, \bibinfo {author} {\bibfnamefont {I.}~\bibnamefont {Bush}},
  \emph {et~al.},\ }\href {http://www.crystal.unito.it/Manuals/crystal14.pdf}
  {\enquote {\bibinfo {title} {Crystal14 user’s manual; university of torino:
  Torino, italy, 2014},}\ }\BibitemShut {NoStop}%
\bibitem [{cry()}]{crystal}%
  \BibitemOpen
  \href {http://www.crystal.unito.it/} {\enquote {\bibinfo {title} {For program
  description, see http://www.crystal.unito.it/},}\ }\BibitemShut {NoStop}%
\bibitem [{\citenamefont {Pakdel}\ \emph {et~al.}(2018)\citenamefont {Pakdel},
  \citenamefont {Pourfath},\ and\ \citenamefont
  {Palacios}}]{pakdel2018implementation}%
  \BibitemOpen
  \bibfield  {author} {\bibinfo {author} {\bibfnamefont {S.}~\bibnamefont
  {Pakdel}}, \bibinfo {author} {\bibfnamefont {M.}~\bibnamefont {Pourfath}}, \
  and\ \bibinfo {author} {\bibfnamefont {J.~J.}\ \bibnamefont {Palacios}},\
  }\href@noop {} {\bibfield  {journal} {\bibinfo  {journal} {Beilstein J.
  Nanotech.}\ }\textbf {\bibinfo {volume} {9}},\ \bibinfo {pages} {1015}
  (\bibinfo {year} {2018})}\BibitemShut {NoStop}%
\bibitem [{\citenamefont {Nomura}\ and\ \citenamefont
  {MacDonald}(2007)}]{PhysRevLett.98.076602}%
  \BibitemOpen
  \bibfield  {author} {\bibinfo {author} {\bibfnamefont {K.}~\bibnamefont
  {Nomura}}\ and\ \bibinfo {author} {\bibfnamefont {A.~H.}\ \bibnamefont
  {MacDonald}},\ }\href@noop {} {\bibfield  {journal} {\bibinfo  {journal}
  {Phys. Rev. Lett.}\ }\textbf {\bibinfo {volume} {98}},\ \bibinfo {pages}
  {076602} (\bibinfo {year} {2007})}\BibitemShut {NoStop}%
\bibitem [{\citenamefont {Pedersen}\ \emph {et~al.}(2001)\citenamefont
  {Pedersen}, \citenamefont {Pedersen},\ and\ \citenamefont
  {Kriestensen}}]{pedersen2001optical}%
  \BibitemOpen
  \bibfield  {author} {\bibinfo {author} {\bibfnamefont {T.~G.}\ \bibnamefont
  {Pedersen}}, \bibinfo {author} {\bibfnamefont {K.}~\bibnamefont {Pedersen}},
  \ and\ \bibinfo {author} {\bibfnamefont {T.~B.}\ \bibnamefont
  {Kriestensen}},\ }\href@noop {} {\bibfield  {journal} {\bibinfo  {journal}
  {Phys. Rev. B}\ }\textbf {\bibinfo {volume} {63}},\ \bibinfo {pages} {201101}
  (\bibinfo {year} {2001})}\BibitemShut {NoStop}%
\end{thebibliography}
%


\end{document}